# Network–Configurations of Dynamic Friction Patterns

H.O. Ghaffari, R.P. Young

*Department of Civil Engineering and Lassonde Institute,*
*University of Toronto, Toronto, ON, Canada*

The complex configurations of dynamic friction patterns–regarding real time contact areas– are transformed into appropriate networks. With this transformation of a system to network space, many properties can be inferred about the structure and dynamics of the system. Here, we analyze the dynamics of static friction, *i.e.* nucleation processes, with respect to "friction networks". We show that networks can successfully capture the crack-like shear ruptures and possible corresponding acoustic features. We found that the fraction of triangles remarkably scales with the detachment fronts. There is a universal power law between nodes' degree and motifs frequency (for triangles, it reads $T(k) \propto k^{\beta}$ ($\beta \approx 2 \pm 0.4$)). We confirmed the obtained universality in aperture-based friction networks. Based on the achieved results, we extracted a possible friction law in terms of network parameters and compared it with the rate and state friction laws. In particular, the evolutions of loops are scaled with power law, indicating the aggregation of cycles around hub nodes. Also, the transition to slow rupture is scaled with the fast variation of local heterogeneity. Furthermore, the motif distributions and modularity space of networks –in terms of within-module degree and participation coefficient–show non-uniform general trends, indicating a universal aspect of energy flow in shear ruptures.

***Keyword:*** *Shear Rupture, Contact areas, Friction Networks, slow rupture*

*Introduction*-The transition from static to dynamic friction with regard to rupture nucleation and precursors is the key feature in the sliding process of frictional interfaces. A basic process in the transition from slip to slid state (stick-slip) includes propagation of detachments fronts where reductions of contact areas yield fast energy flow in the rupture tip. Initiation of detachment fronts is followed by the emission of acoustic waves. Detachment fronts (front-like ruptures) are the wave-like fronts that are formed during the local and global fast change of contact areas, crossed or arrested through the interface [1-3]. By employing recent advancements in data acquisitions systems, laboratory experiments reveal three rupture modes [1-7]: slow ruptures, sub-Rayleigh ruptures (well known as regular earthquakes) and super-shear ruptures. The formation, transition and arresting of shear rupture modes–as well as detachment fronts which are mostly sub-Rayleigh forms– are the most elusive problems in terms of loading configurations and the geometrical complexity of frictional interfaces. Although different numerical and analytical methods investigated several aspects of regular and super-shear ruptures [8-10], the inherent complexities embedded in fault surfaces and the possible collective deformation of an interface's elements necessitates further detailed analysis of shear ruptures. Recent experimental observations [1-5] show a clear pattern of slow ruptures, as well as the transition of regular ruptures to slow fronts; slow ruptures to sub-Rayleigh fronts; or the arresting of the rupture fronts.

Very recently, two approaches with respect to laboratory observations of slow ruptures have been proposed [11-13]. The first involves the continuous formulation of an extended spring-block model presented in [14]. It considers the plasticity of contact areas with proper Heaviside function. It also predicts smaller localization of energy in slow ruptures. The second approach [13] uses rate and state friction laws (with Runia's friction laws [15] for evolution of the state parameter). By employing proper values of coefficient and numerical modeling, it was postulated that the emergence of slow ruptures results from slow creep fronts [13]. From another perspective, careful investigation of the correlation patterns of particles in sheared materials – numerical and experimental evidence [16-17] – showed the emergence of anisotropic long-range correlated patterns during the deformation of the sample. In this study, we investigate the possible dependency of shear rupture transitions with correlation patterns. To characterize the similarity patterns of real-time friction patterns, we map the interface's elements into the proper networks. With this transformation, the networks' parameters are related to the state variable in friction laws, such that frictional interfaces can be modeled in terms of networks' evolution. Remarkably, analysis of different real-time contact measures showed rupture transitions strongly scaled with the motif evolution of networks as well as the fraction of triangles (clustering coefficient). Furthermore, we found emerged assortative networks show a unique power law scaling in terms of global and local similarities. Interestingly, irrespective of rupture speed, similar trends in motifs' ranks are observed. Our findings about the modularity of friction-networks with respect to within-module degree (Z) and participation coefficient (P) indicate that evolvable frictional interfaces occupy certain regions of the modularity space.

***Materials and Methods***- Our data set includes developments of real-time contact areas in recent slip-slid friction experiments on transparent interfaces [1, 3-4]. Generally, the experiments involve shearing two transparent blocks on each other while a uniform normal loading is on the top block and a tangential force is used in the trailing edge. Recordings of the real-time relative contact areas mostly are based on a 1D assumption of interface dimension .It considers average - optical intensity with employing a laser ray through the interface [1, 4]. However, there are limited cases of 2D measures of relative contact areas corresponding to intensity of laser-light. We use a correlation measure over 2D relative contact patterns. Furthermore, for 1D interface, we employ a truncated norm and standard-delay-coordinate embedding method in analysis of time-series to compare the similarity of the patterns. To set up a non-directed network over 2D contact areas in a certain time step, we considered each patch of measured contact areas perpendicular to shear direction as a node –Fig.1a. Each profile has N pixels where each pixel shows the relative-contact area of that cell. Then, we define correlations in the profiles by using:

$$C_{ij} = \frac{\sum_{l=1}^{N} [A_i(l) - <A_i>].[A_j(l) - <A_j>]}{\sqrt{\sum_{l=1}^{N} [A_i(l) - <A_i>]^2} . \sqrt{\sum_{l=1}^{N} [A_j(l) - <A_j>]^2}} \qquad (1)$$

where $A_i(l)$ is $i^{th}$ profile with $1 \leq l \leq N$. Obviously, we could use real-time deformation of contacts area (deformation in $z$ direction) or generally shear displacement components- $\Delta u = (\Delta x, 0, \Delta z)$ where we assumed the interface does not have any shear component in $y$-direction. We may see each profile as a separated cycle of a spatial series (i.e., collection of cycles in $x$-direction). To map the obtained series, we define each patch as a node. To make an edge between two nodes, relative-high correlated profiles are connected ($C_{ij} \geq r_c$) with non-direct links. To choose $r_c$, we use a nearly stable region in rate of betweenness centrality (B.C) - $r_c$ space which is in analogy with minimum value in the rate of edges density [18-19]. The later method has been used successfully in analyzing time-series patterns in network space [20]. We notice finding a nearly stable region in $B.C$-$r_c$ space satisfies dominant structures of contact patterns. Despite the aforementioned case, most of the recorded photos -at the certain time- had insignificant dimension in $z$ direction (<0.2 mm-6mm). Then, naturally with respect to the apparatus, the observed patterns were the longitudinal contact zones. In other words, 2D imaging drastically slows down monitoring in the propagation direction. As a result the measurements were generally 1D, while the integration over the $z$ direction is performed by optical means [2,4]. To build networks over 1D spatial contact areas, we use two different methods. The first method is comparing "closeness" of contact areas; *i.e., if* $| A(x_i, t) - A(x_j, t) | < \xi \rightarrow a_{ij} = 1$ where $a_{ij}$ is the component of connectivity matrix. We use a similar procedure ($B.C$-$\xi$ space) to select the threshold level.

To proceed, we use several characteristics of networks. Each node is characterized by its degree $k_i$ and the clustering coefficient. Clustering coefficient (as a fraction of triangles) is $C_i$ defined as $C_i = \dfrac{2T_i}{k_i(k_i - 1)}$ where $T_i$ is the number of links among the neighbors of node $i$. Then, a node with $k$ links participates on $T(k)$ triangles. Furthermore, based on the role of a node in the modules of network, each node is assigned to its within-module degree ($Z$) and its participation coefficient ($P$). High values of $Z$ indicate how well-connected a node is to other nodes in the same module, and $P$ is a measure of well-distribution of the node's links among different modules []. To determine modularity and partition of the nodes into modules, the modularity $M$ (i.e., objective function) is defined as [22]:

$$M = \sum_{s=1}^{N_M} \Big[ \frac{l_s}{L} - \Big( \frac{d_s}{2L} \Big)^2 \Big], \qquad (2)$$

in which $N_M$ is the number of modules (clusters) , $L = \dfrac{1}{2}\sum_i k_i$ , $l_s$ is the number of links in module and $d_s = \sum_i k_i^s$ (the sum of nodes degrees in module $s$). Using an optimization algorithm (here we use Louvain algorithm [23]), the cluster with maximum modularity is detected.

To describe the correlation of a node with the degree of neighboring nodes, assortatitive mixing index is used:

$$r_k = \frac{<j_l k_l> - <k_i>^2}{<k_i^2> - <k_i>^2} \qquad (3)$$

where it shows the Pearson correlation coefficient between degrees $(j_l, k_l)$ and $<\bullet>$ denotes average over the number of links in the network.

***Results and Discussion***- We start with 2D interfaces, monitored at discrete time steps [0, 0.4, 0.75, 1, 1.2, and 1.4] *ms* [1]. Transferring X-patches (then perpendicular to shear direction) to networks and plotting clustering coefficient revealed three distinct patterns of rupture evolution (Fig1.b) which are comparable with the previous results [1-2]. The three patterns correspond with sub-Rayleigh (1 and 3 in Fig1.b) and slow rupture (2). In other words, the movement of rupture tip is followed by fast variation of clustering coefficient. Furthermore, with considering 3 points cycles (*T*-triangle loops) versus node's degree from 0.4-1.4 ms show a power law scaling (Fig1.c):

$$T(k) \sim k^{\beta}, \qquad (4)$$

where the best fit for collapsed data set reads $\beta \approx 2.1 \pm 0.4$ (we call coupling coefficient of local and global structures). Similar scaling law is obtained for other types of sub-graphs. Peculiarly, Euclidean distance yields the same scaling law. Thus, our analysis of aperture patterns in discrete slip measurements (over 20mm shear displacements) in rock samples reveals the same scaling law with a very close coupling coefficient [18]. With some mathematical analysis, one can show adding $m$ edges increase the number of loops with $\beta^2 m^{\beta}$ which indicates a very congested structure of global and local subgraphs during shear rupture. Also, we notice $C(k) \sim 2k^{\beta-2}$ so that for $\beta < 2$, a hierarchical structure is predicted [25]. Now, we assume variation of clustering coefficient is proportional with variation of shear stress, i.e., $\frac{\partial f_s}{\partial t} \sim -(\frac{\partial C(k)}{\partial t})$. Consequentially, we obtain $\frac{\partial f_s}{\partial t} \sim k^{\beta-2}(2-\beta)\frac{\partial \ln k(t)}{\partial t}$ and in terms of loops it reads $\frac{\partial f_s}{\partial t} \sim 2(2-\beta)T(k)^{(1-\frac{3}{\beta})}$. With $\varepsilon \equiv k^{\beta-2}(2-\beta)$ we obtain $f_s \sim \varepsilon \ln k(t)$. Based on figure 1(b), we also estimate the speed of rupture fronts around the rupture zone with: $\frac{\partial^2 C(x,t)}{\partial x \partial t} \sim 1/v_{front}$ which states the temporal-spatial gradient of the fraction of triangles correlates with the inverse of the speed of rupture fronts.

Next, we look at the distribution of loops (Fig1.e). Irrespective of the different type of rupture modes in the considered time interval, a power law satisfies the obtained distributions:

$$P(T(k)) \sim T^{-\gamma}, \qquad (5)$$

in which $\gamma \approx 2.01$. The power law nature of loops shows aggregation of cycles around "hubs". In other words, during rupture evolution, loops (and generally subgraphs with loops) are not distributed uniformly. In Fig1.f we show friction-networks are assortative, i.e., nodes tend to connect to vertices with a similar degree. The variation of assortativity will be discussed in our analysis of 1D friction patterns. Let transfer (5) into a "hubnness" model (as well as Barabasi-Albert model [26]), where we assume hub nodes tend to absorb nodes with rich loops rather than poor nodes. Then, it leads to:

$$\frac{\partial T_i(k)}{\partial t} \sim m\frac{T_i}{\sum T_j}, \qquad (6)$$

in which $m$ is a coefficient of growth (or decay). Plugging (4) in (6) yields:

. $$\frac{\partial k_i}{\partial t} \sim \frac{m}{\beta}\frac{k_i}{\sum k_j^{\beta}} \qquad (7)$$

For $\beta = 1$ the model yields scale-free networks. To complete our analysis, we write the state parameter of state and rate friction law [29] as a linear combination of local and global structures. We consider a simplified form of the standard equation for the friction with assumption of nearly constant sliding velocity: $\mu_s \sim \ln\frac{\theta}{D_c}$ in which $\theta$ is the variable describing the interface state and $D_c$ is the characteristic length for the evolution of $\theta$. The state variable carries information about the whole population of asperities [30]. Commonly used empirical laws for evolution of state variable are Ruina's laws for ageing and slipping states [15]. For slip law, it reads: $\frac{\partial \theta}{\partial t} \sim -\frac{\theta}{D_c}\ln\frac{\theta}{D_c}$. Let transfer state variable in networks space as follows:

$$\frac{\partial \theta_i(t)}{\partial t} = a\frac{\partial k_i}{\partial t} + b\frac{\partial T_i}{\partial t} \qquad (8)$$

which indicates the evolution of the state variable is a superposition of local and global friction networks. We eventually obtain:

$$\frac{\partial \theta_i(t)}{\partial t} = \frac{\partial k_i}{\partial t}(a + b\beta k_i^{\beta-1}) \qquad (9)$$

Assuming $\beta = 1, \frac{\partial k_i}{\partial t} > 0$ and $a < 0$ indicates a decaying model for the state parameter in terms of attacking to hubs. Plugging (7) into (9) and assuming $\beta = 2$ leads to:

$$\frac{\partial \theta_i(t)}{\partial t} = \frac{2bmk_i^2}{\sum k_j^2} + \frac{amk_i}{\sum k_j^2} \qquad (10)$$

which shows a complex non-linear nature of state parameter. Further developments of (9) and (10) can be done by using non-linear growth models as well as rate equations [27]. Notably, the first part at the left hand of Eq.10 shows friction networks follow a gel-like state, in which a condensation of nodes occurs [27-28]. In this case (or when $\frac{a}{b} \to 0$), a single node is connected to almost all nodes in the friction network. This behavior is biased with the second term if the weight of global evolution is significant. Considering the spatial gradient of the state variable's rate and recalling the relation of rupture speed and clustering coefficient requires (writing Eq.8 in terms of $k$ and $C$ and plugging $\frac{\partial^2 C(x,t)}{\partial x \partial t} \sim 1/v_{front}$): $1/v_{front} \sim \chi(k)\frac{\partial^2 \theta(x,t)}{\partial x \partial t}$ where $\chi(k)$ is a function of node's degree and coupling coefficient. To confirm this relation and using the methodology used in [11-12], we estimate front's speed. Assuming $\frac{\partial \bullet}{\partial t} = -v_{front}\frac{\partial \bullet}{\partial x}$ and recalling Ruina's law with the assumption of unity rate of displacement, $1 - \frac{\theta}{D_c}$; then we obtain

$v_{front}\frac{\partial^2 \theta(x,t)}{\partial x \partial t} + \frac{\partial v_{front}}{\partial t} = \frac{-1}{D_c}$ which is comparable with our prediction of the front's speed. It is noteworthy $\frac{\partial v_{front}}{\partial t}$ indicates the inertia of rupture's front. In the following, we also discuss the characteristic length in terms of the modularity parameters of friction networks.

Next, we plot sub-graphs (*i.e.,* motifs) distributions (Fig.1d) which indicate a super-family phenomenon. A similar trend in all the rupture speeds indicates universality in energy flow in frictional interfaces which is characterized by friction laws. We examined this universality by measuring contacts and aperture in terms of discrete displacement intervals and confirmed the achieved results [18]. Remarkably, transition from sub-Rayleigh to slow rupture is correlated with a distinct spike in all types of the 4-points sub-graphs. Considering the variations of <*k*> at the monitored interval ($10 \le k \le 30$) and the abrupt change in loops which is in order of one magnitude, we conclude that the fluctuation of coupling coefficient induces remarkable growth of sub-growth. One can confirm increasing sub-graphs is consistent our formulation and observation (for 1D-friction patterns), and consequently increases $\beta$ which leads to rapid increase of loops.

Following 1-D patterns of contact areas, we map 1D-net contact areas into friction networks by using closeness metric (Figure 2). According to [3] – and figure 2b-, the evolution of contact areas follows three remarkable evolutionary stages as follows: 1) detachment phase, 2) rapid slip, and 3) slow slip. We notice in this case, the first phase accompanies around 10% of contact area variation while phase 2 represents 2-3% of the changes in contacts. Notably, the evolution of pure contact areas does not reveal any more information on rupture speed or other possible mechanism behind the fracture evolution. However, the corresponding friction networks parameters, such as assortativity index and maximum modularity (Q) [31], indicate more features of the rupture speed. Each front passage encoded as the remarkable spikes in assortativity index or maximum modularity (Figure 2b, f). We also confirmed the coupling coefficient dramatically drops due to the front's passage. Based on our formulation, the rapid drop of coupling coefficient induces fast variation of shear stress.

Moreover, the maximum variation s of assortativity and modularity at the first phase are ~50% and ~30%, respectively. A significant spike in transition to the phase (1) is followed by a nearly large stationary value of assortativity, where the transition time is comparable to the propagation of sub-Rayleigh's front. Generally, our observation showed rapid sub-Rayleigh fronts encoded as sudden spikes in assortativity index, comparable with the radiated energy and acoustic emissions. Remarkably, the slow rupture stage is generally represented by monotonic growth of modularity while the variation of assortativity occurs in the two distinct categories: stable and non-stable slow rupture. Stability in assortativity occurs in two different stages; one is before sliding and another one takes place in net-movement of the interface. However, it seems the mechanisms of the similar stable-assortative stages are different. The first stationary interval accompanies the growth of modularity while the continuous decaying of links happens. The second interval, which occurs with the same growth of assortativity (>20%), shows a decreasing trend in modularity. We concluded after a rapid-strong sub-Rayleigh front that a silent period is encoded in the system, followed by a fast transition to next stage. Indeed, transition to phase3 occurs with ~25% fast-increment of assortativity index while the variation of the contact areas is only less than 4%. In other words, during this small time-interval (~60-100μs), what controls the behavior of the interface is the pattern (shape) of contacts rather than pure contact areas where we guess the central rich-hubs drive the system. It immediately follows a relatively long stationary assortativity, indicating a stable mechanical characteristic of the interface (possible low sliding rate [1]).

With respect to our observations, we hypothesize stable and unstable-slow ruptures carry high and small localization, respectively, while sub-Rayleigh ruptures hold high localization in terms of the evolution's rate of the assortativity. Obviously, irrespective of slow rupture modes, the growth of modularity is its unique characteristic. Phase 3, clearly is characterized by a long term of nearly stability in assortativity, net-contact areas, and node's degree (with a small ~3% growth in modularity, clustering coefficient are observed). We have confirmed theses results with three more data sets. It is worth mentioning that the maximum eigenvalues of the Laplacian matrix show approximately the same trend of degree correlation (Figure.2g).

Mapping the nodes of friction networks in modularity space, defined by within module index (Z) and participation coefficient (P) revealed collapsing of all nodes in a certain range of P and Z (Figure2C). We confirmed the universality of observed patterns in P-Z space across different case studies: either real time contacts (1 or 2 D) or aperture friction networks. Following [21], we divide the P-Z space into 7 sub-regions: $R_1, ..., R_7$ which is based on the role of non-hubs and hubs nodes. The general categorization of P-Z parameter space is as follows:

$Non-hubs : \{R_1(P \le 0.05), R_2(0.05 < P \le 0.62), R_3(0.62 < P \le 0.80), R_4(P > 0.80)\} \cap \{Z < 2.5\}$,

$Hubs : \{R_5(P \le 0.3), R_6(0.3 < P \le 0.75), R_7(P > 0.75)\} \cap \{Z \ge 2.5\}$,

where each region nominates a unique characteristic of rich or poor nodes with respect to other modules. Our results for 8 case studies show that the probability of finding a profile in $R_4$ and $R_7$ is very unlikely, as most congestion of profiles happens in $R_1$, $R_2$ and $R_3$. A few nodes occupy $R_5$, indicating the role of hub nodes with the majority of connections within their module. We noticed a general negative fit of collapsed nodes in P-Z space. Amazingly, plotting the evolution of within module index in *x-t*-Z(*x,t*) shows a periodic nature (Figure 2*i*). We obtained the periodic property of Z*(x,t)* in other real-time contact areas and aperture-friction networks. We believe the periodicity of within-module degree and participation coefficient -with respect to nearly the invariant nature of spatial-periodicity- are related somehow to the characteristic length ($D_c$) of the interface. Interestingly, in this parameter space, a clear super-shear front is revealed. Further studies will highlight the network-signatures of super-shear fronts (rupture). As well as the previous case, we can develop the evolution of the state parameters with respect to modularity parameters:

$$\frac{\partial \theta}{\partial t} = c_z \frac{\partial Z}{\partial t} + c_P \frac{\partial P}{\partial t}$$

$$\text{Assuming} \quad \frac{\partial^2 Z}{\partial x \partial t} \sim 1/v_{front} \quad then \quad \frac{\partial \theta}{\partial t} = \frac{\partial Z}{\partial t}(c_z + c_P \frac{\partial P}{\partial Z})$$ (11)

$$\frac{\partial^2 \theta}{\partial x \partial t} \sim v_{front}^{-1}(c_z + c_P \frac{\partial P}{\partial Z}) + c_P \frac{\partial Z}{\partial t}\frac{\partial^2 P}{\partial x \partial Z}$$

in which $c_z$ and $c_P$ are appropriate constant variables (i.e., weights of modularity parameters) .

We estimate the slope of upper (or lower) bond of P-Z space with a constant gradient: $\frac{\partial P}{\partial Z} \approx -\delta$ .

Hence, it leads to: $(\frac{1}{c_z - \delta c_P})\frac{\partial^2 \theta}{\partial x \partial t} \sim 1/v_{front}$ ,indicating increasing $\delta$ leads to decreasing rupture's speed. We conclude that fast fronts induce (or generally scale with) out-of the modules developments and slow ruptures scale with in-modules evolution.

***Conclusions***-As a conclusion to our study, we introduced friction networks over dynamics of different real time contact areas. Based on our solid observations, we formulated a probabilistic frame for the evolution of the state variable in terms of friction networks. Moreover, we confirmed that slow ruptures generally hold small localization, while regular ruptures carry a high level of energy localization. We also introduced two new universalities with respect to the evolution of dry frictional interfaces: the scaling of local and global characteristics and the occupation of certain regions of modularity parameter space. Our results showed how the relatively highly correlated "elements" of an interface can reveal more features of the underlying dynamics. We proposed that assortativity as an index to correlation of node's degree can completely uncover acoustic features of the interfaces. Our formulation can be coupled with elasto-dynamic equations to complete our understanding of the interface's more realistic features.

• We would like to acknowledge and thank Prof. **J. Fineberg** (The Racah Institute of Physics, Hebrew University of Jerusalem) and **S. Maegawa** (Graduate School of Environment and Information Sciences, Yokohama National University, Japan) who provided the data set employed in this study.

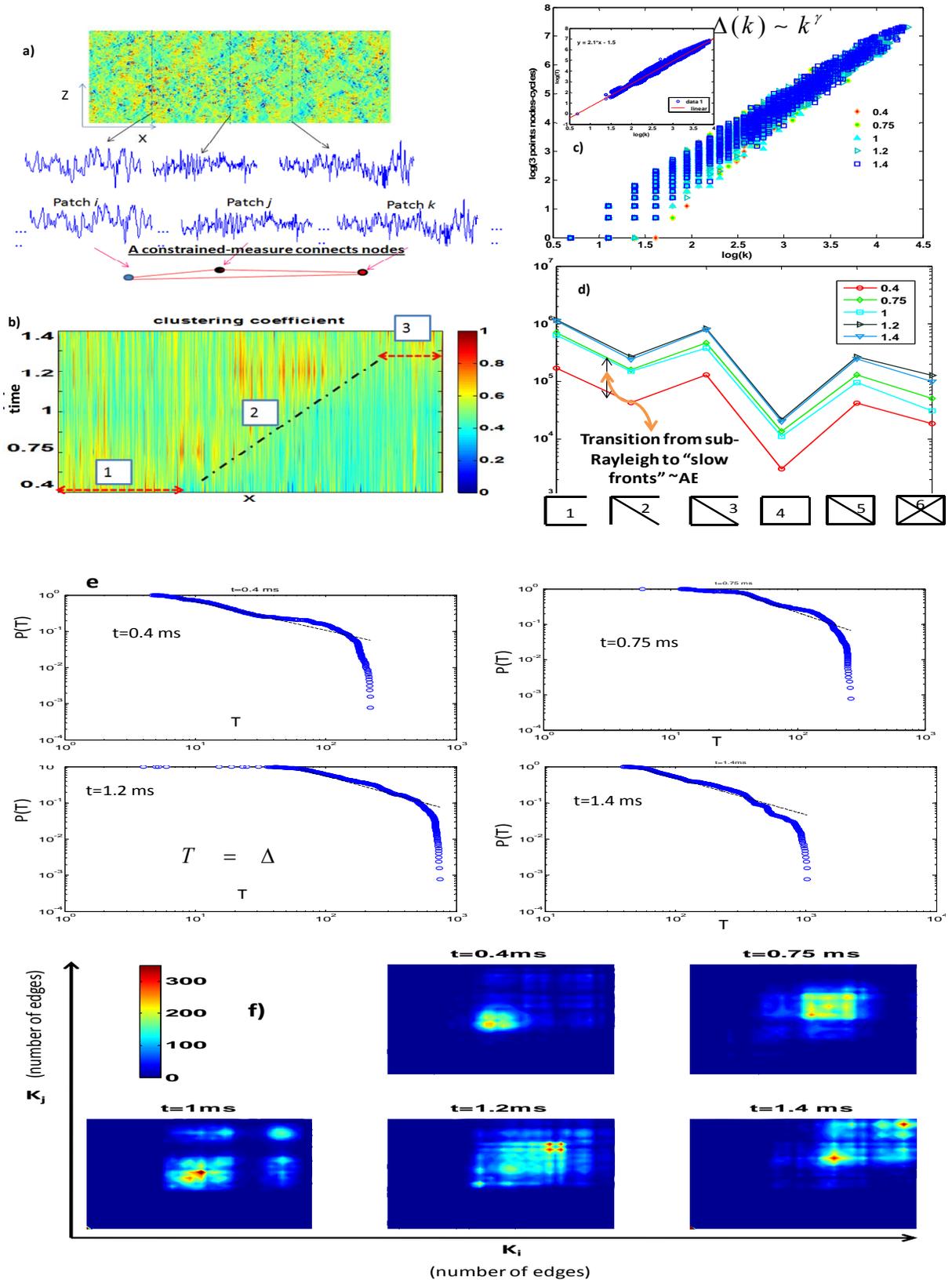

a)

Patch i   Patch j   Patch k

A constrained-measure connects nodes

b) clustering coefficient

c) $\Delta(k) \sim k^{\gamma}$

log2 points nodes-cycles

log(k)

* 0.4
* 0.75
△ 1
▽ 1.2
□ 1.4

d)

Transition from sub-Rayleigh to "slow fronts" ~AE

1   2   3   4   5   6

e)

t=0.4 ms

t=0.75 ms

t=1.2 ms     $T = \Delta$

t=1.4 ms

f)

$\kappa_j$ (number of edges)

$\kappa_i$ (number of edges)

t=0.4ms   t=0.75 ms

t=1ms   t=1.2ms   t=1.4 ms

Figure 1. (a) A typical example of transferring contact areas into networks ,(b) mapping the dynamic relative contact areas in 2D [1] to networks space an  plotting clustering coefficient as a fraction of triangles reveals the relatively precise rupture speed  :three distinctive rupture  speeds is compatible with mean contact area ;1 , 3 corresponds with sub-Rayleigh rupture and 2  is slow rupture ; (c) scaling of triangles in obtained networks with  number of similar profiles (node's degree)  ,expressed with power law relation  with universal exponent  (~2) ;(d) distribution of motifs of networks over different time steps shows a non-uniform /universal distribution ;transition from rupture(1) to rupture (2) occurs suddenly which   is related to regular acoustic emission (high frequency waveform) ;(e) distribution of triangles shows power-law distribution –This proves loops congests around "hubs" and This  property drives  frictional interfaces ;(f) Joint degree  distribution as a two-point correlation of edges shows the obtained networks are assortative : rich-hubs attaching to rich nodes . Furthermore, motif distributions have a similar trend, indicates a universal aspect of energy flow in shear ruptures.

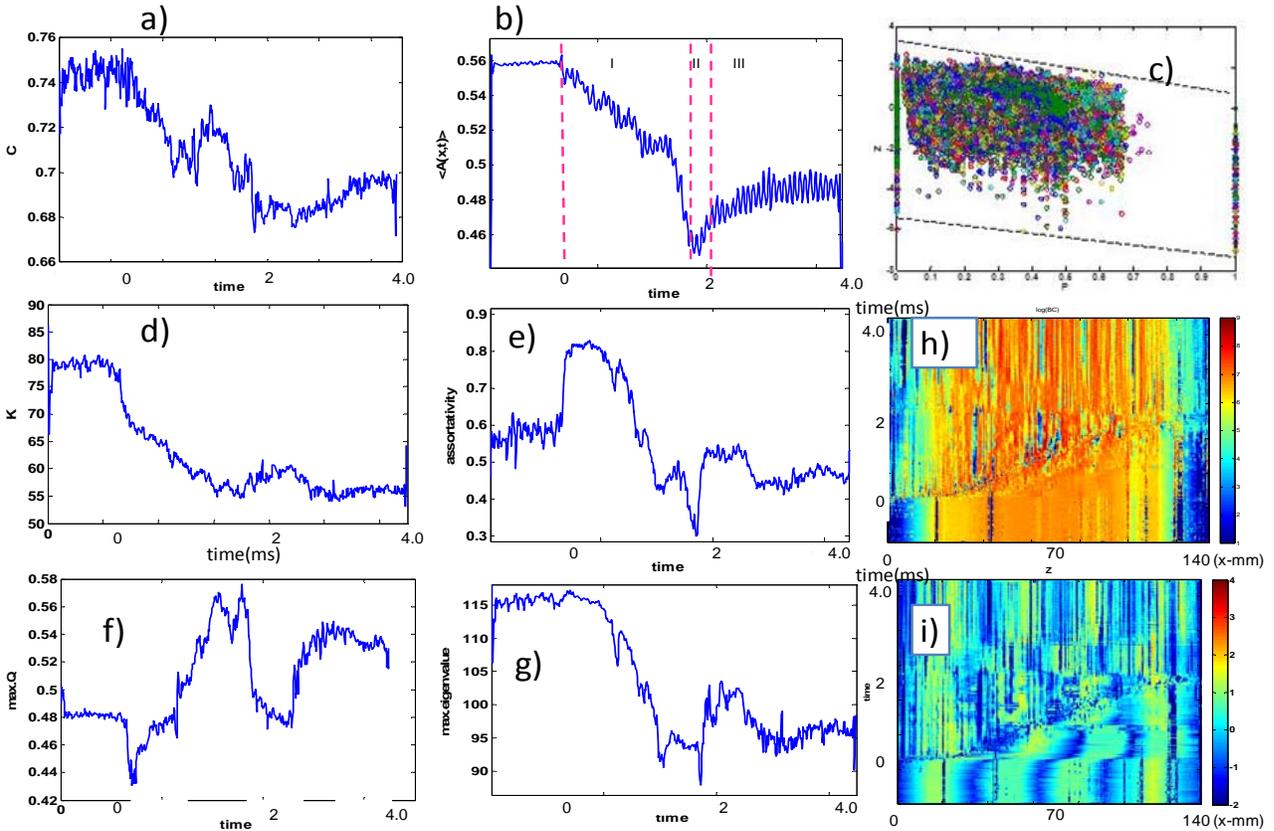

Figure 2. Transferring 1D contact areas [1] to networks with spatial constrained metric  ;(a) clustering coefficient versus time (~4ms),(b) real-time contact areas nearly show three distinct regimes : slip stage ;aging/sliding (c) each element of interface-network is mapped into modularity space (participation factor of each edge and within module degree)-For different case studies we found elements of interfaces occupy  certain regions of P-Z space  ;(d) node's degree versus time (~4ms) nearly follow contact areas variation-This yields we may write state parameter in terms of networks variables  ;(e) assortativity coefficient versus time (~4ms) shows all of networks are assortative ; transition to new rupture speed dramatically change assortativity- In Transition to aging step ,as if  real time contacts are changing less than 3-5% how assortativity changes 15-25%;(f) maximized Q as an index to modularity versus time  ;(g) maximum eigenvalues of Laplacian of node's degree against time indicate possible synchronization of elements of frictional interfaces;(h) Betweenness Centrality(B.C) –as a  measure of a node's centrality in a network-indicate clear rupture transitions and possible periodic nature of interface (characteristic length in friction law);Here we have plotted natural logarithmic variation of B.C  (i) variation of within module degree (Z) precisely shows intersonic fronts and details of fronts ;periodic nature of modules clearly affected by detachment fronts.